\documentclass[letterpaper,journal,comsoc]{IEEEtran}

\usepackage{amsmath,amssymb,amsfonts}
\usepackage{booktabs}
\usepackage{graphicx}
\usepackage{cite}
\usepackage{url}
\usepackage[hidelinks]{hyperref}
\hypersetup{
  pdftitle={Sense Once, Serve Many: Common-Trace Factorized Constrained PPO for Online Sensing-Session Consolidation in Multi-Tenant ISAC Networks},
  pdfauthor={Dang-Dung Vu}
}

\begin{document}

\title{Sense Once, Serve Many: Common-Trace Factorized Constrained PPO for Online Sensing-Session Consolidation in Multi-Tenant ISAC Networks}

\author{Dang-Dung Vu%
\thanks{(Corresponding author: Dang-Dung Vu.)
The author is with the Faculty of Information Technology,
VNU University of Engineering and Technology (VNU-UET),
Hanoi 100000, Vietnam
(e-mail: dungvd21105@gmail.com).}
}

\maketitle

\begin{abstract}
Integrated sensing and communication (ISAC) networks can serve compatible requests through shared sensing sessions, but online consolidation couples admission, reuse, resource-profile selection, sensing service-level agreements (SLAs), communication quality of service (QoS), and future commitments. We formulate sensing-session consolidation as a constrained Markov decision process and propose Common-Trace Factorized Constrained Proximal Policy Optimization (CT-PPO). During training, stochastic policy replicas share the same exogenous workload trace; leave-one-out discounted Monte Carlo return contrasts provide reward credit to applicable action factors, while constraint credit remains factor-specific. Across five training seeds and matched workloads, CT-PPO exceeds matched Joint-Credit PPO (JC-PPO) by $0.934$ in macro return (95\% confidence interval $[0.702,\,1.164]$), reduces sensing-resource cost by $6.277$, and increases accepted requests per created session by $0.0806$. A four-way ablation shows that replacing JC-PPO's joint-ratio surrogate with a factor-wise surrogate alone yields no detectable macro-return gain, whereas adding common-trace reward credit produces the dominant improvement. Without retraining, CT-PPO retains a positive return advantage at low, nominal, and high tested arrival loads. Deployment uses public observations and hard feasibility masks; CT-PPO's extra parameters are confined to training-side prefix critics, its deployed encoder--actor footprint matches JC-PPO, and actor-only batch-one CPU latency is effectively unchanged.
\end{abstract}


\section{Introduction}
\label{sec:introduction}

Integrated sensing and communication (ISAC) is moving beyond joint radar--communication design toward networked systems in which sensing is requested, orchestrated, and delivered as a service~\cite{liu2020joint,dong2023saas}. Heterogeneous applications may demand detection, localization, or tracking with different spatial, temporal, and accuracy requirements while sharing finite radio resources with communication traffic. Sensing-as-a-service and cross-layer architectures provide the corresponding request, control, resource-management, and service-delivery abstractions~\cite{dong2024saas,wymeersch2025crosslayer}, while architectural work explicitly models sensing-service consumers and sensing-result exposure~\cite{etsi2026isc003}. These service abstractions leave an online consolidation decision: as requests arrive, which should reuse an active sensing session, which should create a new one, and when should service be deferred or rejected?

Online consolidation is stateful because a merge must preserve the service already committed to existing members. Compatibility depends jointly on target identity, spatial coverage, task capability, sensing quality, freshness, authorization, timing, and deterministic resource reservations. The orchestrator is therefore an intelligent network-control entity that must choose among \textsc{Merge}, \textsc{Create}, \textsc{Defer}, and \textsc{Reject}, together with the applicable sensing-resource profile. Each decision changes not only current sensing occupancy but also future session commitments; every scheduled sensing update consumes bandwidth and power that would otherwise remain available to communication users. We formulate this coupled service--resource problem as a constrained Markov decision process (CMDP), separating hard physical/contractual feasibility from long-term sensing service-level agreement (SLA) residuals and communication quality-of-service (QoS) residuals, and from the completed-service-versus-sensing-cost objective. The learned controller operates only over these service/session decisions and discrete sensing profiles: communication scheduling is fixed, while waveform, beamforming, precoding, and other continuous physical-layer variables are not policy decisions.

The control policy is itself structured: it selects an action type and, conditionally, a destination session and resource profile. Factorizing the action distribution does not remove the training difficulty created by stochastic arrivals and physical processes, whose variation is superimposed on the return of the policy's composite decisions. We address this with \emph{Common-Trace Factorized Constrained Proximal Policy Optimization (CT-PPO)}. CT-PPO combines a permutation-invariant Set encoder with a masked factorized actor. During training, stochastic policy replicas are exposed to the same exogenous workload trace---the same request arrivals and physical-process realizations---and leave-one-out discounted Monte Carlo return contrasts aligned by physical slot provide a common reward advantage to the applicable action factors. Constraint advantages remain specific to the applicable action factor within the primal--dual PPO update. Common-trace replication and auxiliary prefix critics are used only during training; validation, external evaluation, and deployment use a single policy trajectory with the current public observation and hard feasibility masks, without future workload information or oracle outcomes.

We evaluate CT-PPO against four deterministic online heuristics, Random Valid, and matched Joint-Credit PPO (JC-PPO), holding the environment, public observation, hard masks, Set representation, action distribution, reward and constraints, optimizer family, training budget, validation rule, and external workloads fixed. Across five seeds and matched external traces, CT-PPO attains the highest mean macro return; relative to JC-PPO it gains $0.934$ (95\% CI $[0.702,\,1.164]$), reduces sensing-resource cost by $6.277$, and raises accepted requests per created session by $0.0806$, while exceeding SLA-Aware Greedy by $1.847$. A controlled four-way ablation attributes the return gain primarily to common-trace reward credit: the factor-wise surrogate alone yields no detectable macro-return improvement, whereas adding common-trace reward credit after controlling for that surrogate change shifts behavior toward greater session reuse and produces the main return gain. The subsequent prefix-critic machinery for factor-specific constraint credit yields no detectable incremental return effect in the evaluated configuration. Without retraining, frozen validation-selected CT-PPO and JC-PPO checkpoints retain a positive CT--JC macro-return difference at low, nominal, and high tested arrival loads for all five training seeds. The additional CT-PPO parameters are confined to training-side prefix critics; the deployed encoder--actor parameter count is identical to JC-PPO, and measured actor-only batch-one CPU inference latency is effectively unchanged.

The main contributions are:
\begin{itemize}
    \item We formulate online multi-tenant sensing-session consolidation as a service-level CMDP with explicit \textsc{Merge}, \textsc{Create}, \textsc{Defer}, and \textsc{Reject} decisions, conditional session/profile selection, heterogeneous sensing SLAs, communication QoS, and future resource commitments.

    \item We develop CT-PPO, a constrained factorized policy-learning method that uses stochastic replicas sharing the same exogenous workload trace and leave-one-out Monte Carlo reward contrasts to control for exogenous workload variation in reward credit, while retaining factor-specific constraint credit without changing the information available to the policy at deployment.

    \item We isolate the policy-update components through a controlled four-way ablation. Replacing the joint-ratio surrogate with a factor-wise surrogate alone does not yield a detectable macro-return improvement; after controlling for this change, common-trace reward credit is the dominant empirical driver of the observed return gain, while the additional prefix-critic machinery for factor-specific constraint credit has no detectable incremental return effect in the evaluated configuration.

    \item We provide controlled five-seed external evaluation against matched JC-PPO and practical online heuristics, linking the return gain to selective session reuse and its stronger effect under clustered arrivals. Frozen-checkpoint load-shift evaluation further shows a positive CT--JC return advantage at all three tested arrival loads, while deployment measurements show that CT-PPO's additional parameters are confined to training-side prefix critics and leave the deployed encoder--actor footprint unchanged.
\end{itemize}

The remainder of the paper reviews related work in Section~\ref{sec:related_work}, formulates the system in Section~\ref{sec:system_model}, presents CT-PPO in Section~\ref{sec:method}, describes the evaluation in Section~\ref{sec:experiments}, and reports results, discussion, and conclusions in Sections~\ref{sec:results}--\ref{sec:conclusion}.

\section{Related Work}
\label{sec:related_work}

\subsection{Network-Level ISAC Services and Orchestration}

ISAC has evolved from joint radar--communication design toward dual-functional and network-level systems that provide sensing alongside communication services~\cite{liu2020joint,liu2022isac}. Sensing-as-a-service work makes this transition explicit by mapping heterogeneous detection, localization, and tracking requirements to task-dependent sensing QoS and joint sensing--communication resource allocation~\cite{dong2023saas,dong2024saas}. Standardization and architectural studies further elevate sensing to a managed network service: 3GPP Release~19 studies ISAC use cases and service requirements, while ETSI GR ISC~003 specifies system- and RAN-level architectural functions for sensing control and result exposure~\cite{threegpp2024isac,etsi2026isc003}. Recent network-level and cross-layer ISAC visions likewise emphasize coordinated sensing, communication, information sharing, and resource orchestration across nodes and protocol layers~\cite{han2025networklevel,wymeersch2025crosslayer}.

Resource management spans joint sensing-service allocation~\cite{dong2023saas}, opportunistic spatiotemporal radio-resource reuse~\cite{li2025reuse}, and progressively unified resource allocation under changing system states~\cite{li2026elastic}. Deep reinforcement learning has also been used for ISAC resource allocation and trajectory planning~\cite{qin2023drlisac}. Outside ISAC, sensor-cloud work aggregates application sensing requests into a consolidated sensing schedule shared by multiple applications~\cite{dinh2016sensorcloud}. ISAC architectural work likewise notes request aggregation into one procedure or session~\cite{dass2024privacy}. These works address resource optimization or request aggregation, but not our online stateful decision of whether each arrival reuses an active sensing session, creates one, defers, or is rejected while preserving heterogeneous member commitments and future reservations.

\subsection{Constrained and Set-Structured Policy Learning}

The sensing-SLA and communication-QoS requirements in our formulation connect directly to constrained reinforcement learning. Constrained policy optimization (CPO) optimizes expected return subject to expected-cost constraints~\cite{achiam2017cpo}; proximal policy optimization (PPO) provides the clipped policy-optimization backbone~\cite{schulman2017ppo}; and primal--dual methods such as proportional--integral--derivative (PID) Lagrangian control adapt constraint multipliers during learning~\cite{stooke2020pid}. In the consolidation CMDP, these long-term residual constraints coexist with deterministic action-validity conditions. Hard masks therefore enforce physical, contractual, temporal, quality, and reservation feasibility, while post-admission sensing SLA and communication QoS remain explicit CMDP residuals optimized by the constrained policy update.

The observation also contains variable-cardinality sets of waiting requests and active sensing sessions. A permutation-invariant Set encoder following the Deep Sets principle provides a natural representation for these collections~\cite{zaheer2017deepsets}. CT-PPO and the matched JC-PPO reference deliberately share the same Set representation, public features, masks, and conditional action distribution. This controlled architecture makes the learned comparison focus on how credit is constructed for the structured policy rather than confounding credit assignment with a different state representation or action interface.

\subsection{Structured Credit Assignment and Matched Rollouts}

Credit assignment becomes more difficult when a policy contains conditional action components and observed return also reflects stochastic events outside the policy's control. Action-dependent factorized baselines reduce policy-gradient variance by conditioning baselines on subsets of a multidimensional action~\cite{wu2018factorized}. In cooperative multi-agent learning, COMA assigns credit through counterfactual action baselines~\cite{foerster2018coma}. Counterfactual credit assignment has also been studied explicitly as a means of separating the effect of an action from external random events and subsequent behavior~\cite{mesnard2021counterfactual}. A complementary variance-reduction principle is to couple stochastic rollouts: the vine estimator in TRPO uses common random numbers when comparing rollouts from a shared state so that differences in estimated action values are less dominated by simulation noise~\cite{schulman2015trpo}.

CT-PPO applies matched randomness at a different granularity suited to online sensing-session consolidation. Stochastic policy replicas share the same complete primitive workload trace, fixing exogenous arrivals and physical processes, while their actions generate different endogenous request and session trajectories. Reward returns are compared at the same physical slot using a leave-one-out peer baseline, and the resulting common matched-trace reward advantage is supplied to the applicable factors of the conditional actor. Constraint credit remains specific to the applicable action factor. Unlike action-marginal baselines, CT-PPO does not enumerate alternative action components; unlike future-conditioned counterfactual credit, it does not condition credit on learned future-event information; and unlike same-state branching rollouts, replicas need not remain in the same endogenous state. The resulting credit construction directly targets exogenous workload variation while preserving the online observation and action interface used at deployment.

\section{System Model and Problem Formulation}
\label{sec:system_model}

\subsection{System and Shared-Session Semantics}

\begin{figure*}[t]
\centering
\includegraphics[width=1.00\textwidth]{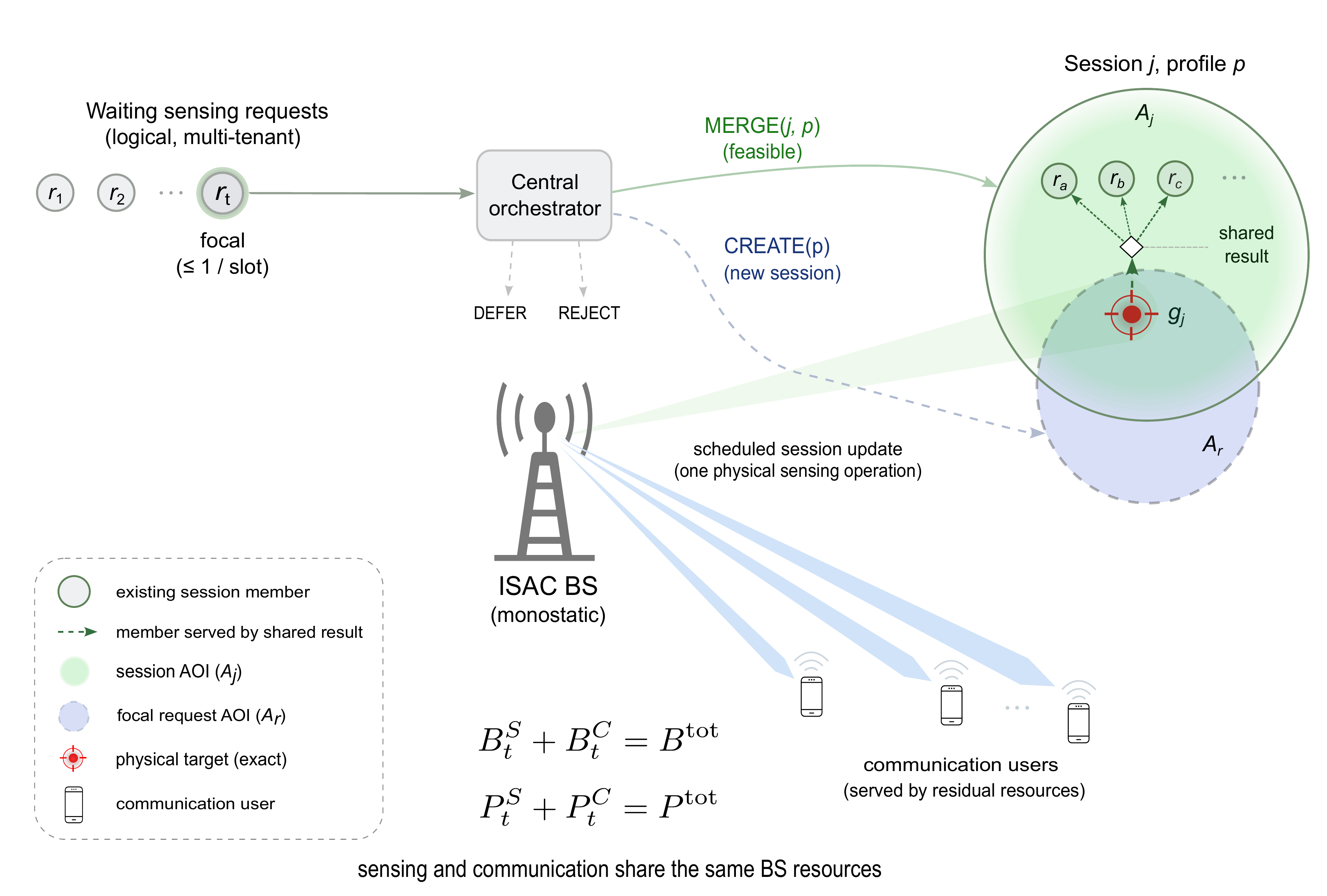}
\caption{Multi-tenant ISAC system and online sensing-session consolidation. At most one focal request is processed per slot. A feasible request may merge into an active sensing session or create a new one; one scheduled session update produces a shared sensing result that is evaluated against each member's own service requirements. Sensing bandwidth and power reduce the residual resources available to communication users.}
\label{fig:system_model}
\end{figure*}

We consider a centralized single-cell ISAC system with sensing tenants $\mathcal L=\{1,\ldots,L\}$, communication users $\mathcal K=\{1,\ldots,K\}$, discrete sensing-resource profiles $\mathcal P$, and active sensing sessions $\mathcal J_t$. One monostatic ISAC base station (BS) shares total bandwidth $B^{\mathrm{tot}}$ and power $P^{\mathrm{tot}}$ over slots $t=0,\ldots,T-1$. After terminal cleanup and the current-slot physical-process and request updates, at most one focal request is exposed for control; it is selected lexicographically by eligibility, arrival, and identifier, independently of value, compatibility, or policy score. The focal action precedes sensing service, residual communication service, and completion/constraint accounting. Figure~\ref{fig:system_model} summarizes the decision and shared-resource coupling.

A request is $x_r=(l_r,a_r,d_r,A_r,g_r,\kappa_r,\theta_r,\Delta_r,v_r,\chi_r)$, denoting tenant, arrival/latest-start slots, disk-shaped area of interest (AOI), exact target, task $\kappa_r\in\{\mathrm{DET},\mathrm{LOC},\mathrm{TRK}\}$ (detection/localization/tracking), quality threshold, maximum valid-output age, completion value, and sharing permission. If admitted at $s_r$, its $H_{\kappa_r}$-slot service ends at $f_r=s_r+H_{\kappa_r}-1<T$. Requests evolve from \textsc{Waiting} to \textsc{Active}, \textsc{Expired}, or \textsc{Rejected}; active requests terminate as \textsc{Completed} or \textsc{Failed}.

Session $j$ fixes AOI $A_j$ and target $g_j$ and stores its member set $\mathcal S_{j,t}$, profile $p_{j,t}$, update calendar, lifetime, output-capability set $\mathcal O_j$ determined by the creator task, and a shared tracking covariance when applicable. The output contracts are $\mathcal O(\mathrm{DET})=\{\mathrm{DET}\}$, $\mathcal O(\mathrm{LOC})=\{\mathrm{DET},\mathrm{LOC}\}$, and $\mathcal O(\mathrm{TRK})=\{\mathrm{DET},\mathrm{LOC},\mathrm{TRK}\}$. A profile $p=(B_p^{\mathrm s},P_p^{\mathrm s},\nu_p)$ specifies sensing bandwidth, power, and update period. Every feasible \textsc{Create} or \textsc{Merge} causes a real current-slot sensing update; subsequent updates recur every $\nu_p$ slots within the session lifetime. One scheduled session update is one physical sensing operation and produces one shared result, which every member evaluates against its own task, threshold, freshness bound, and service interval; no member-specific physical sensing realization is generated inside a shared session.

\subsection{Shared Resources and Service Quality}

Let $\mathcal J_t^{\mathrm{upd}}$ be the sessions updating at slot $t$. Sensing occupancy and residual communication resources obey
\begin{align}
B_t^{\mathrm s}&=\sum_{j\in\mathcal J_t^{\mathrm{upd}}}B_{p_{j,t}}^{\mathrm s}, &
P_t^{\mathrm s}&=\sum_{j\in\mathcal J_t^{\mathrm{upd}}}P_{p_{j,t}}^{\mathrm s},\nonumber\\
B_t^{\mathrm c}&=B^{\mathrm{tot}}-B_t^{\mathrm s}, &
P_t^{\mathrm c}&=P^{\mathrm{tot}}-P_t^{\mathrm s},
\label{eq:resource_coupling}
\end{align}
with $B_t^{\mathrm s}\le B^{\mathrm{tot}}$ and $P_t^{\mathrm s}\le P^{\mathrm{tot}}$. For the $K_t^+$ positive-demand users, the fixed scheduler sets $B_{k,t}^{\mathrm c}=B_t^{\mathrm c}/K_t^+$ and $P_{k,t}^{\mathrm c}=P_t^{\mathrm c}/K_t^+$, with zero allocation when $D_{k,t}=0$; $K_t^+$ is the positive-demand user count and communication scheduling is not an RL action. With demand $D_{k,t}$, gain $g_{k,t}^{\mathrm c}$, noise power spectral density $N_0$, noise factor $F_{\mathrm c}$, and implementation gap $\Gamma_{\mathrm c}$, the served rate is
\begin{equation}
R_{k,t}^{\mathrm c}=\min\!\left\{D_{k,t},\;B_{k,t}^{\mathrm c}\log_2\!\left(1+\frac{P_{k,t}^{\mathrm c}g_{k,t}^{\mathrm c}}{N_0F_{\mathrm c}B_{k,t}^{\mathrm c}\Gamma_{\mathrm c}}\right)\right\}.
\label{eq:served_comm_rate}
\end{equation}

For sensing, $G_{\mathrm s}$, $\lambda_{\mathrm c}$, $L_{\mathrm s}$, and $F_{\mathrm s}$ denote front-end gain, carrier wavelength, system loss, and noise factor, while $\sigma_{g_j,t}$, $d_{j,t}^{\mathrm s}$, $\Xi_{g_j,t}^{\mathrm s}$, and $|h_{g_j,t}^{\mathrm s}|^2$ denote target RCS, BS--target distance, shadowing gain, and fading power gain. The echo gain $g_{j,t}^{\mathrm s}$, sensing SNR $\gamma_{j,p,t}^{\mathrm s}$, and detection probability $P_{\mathrm D,j}$ are
\begin{align}
g_{j,t}^{\mathrm s}
&=\frac{G_{\mathrm s}\lambda_{\mathrm c}^{2}\sigma_{g_j,t}}
{(4\pi)^3(d_{j,t}^{\mathrm s})^4L_{\mathrm s}}
\Xi_{g_j,t}^{\mathrm s}|h_{g_j,t}^{\mathrm s}|^2,\quad
\gamma_{j,p,t}^{\mathrm s}=\frac{P_p^{\mathrm s}g_{j,t}^{\mathrm s}}{N_0F_{\mathrm s}B_p^{\mathrm s}},\nonumber\\
P_{\mathrm D,j}(p,t)
&=1-F_{\chi'^2_2(2\gamma_{j,p,t}^{\mathrm s})}(-2\ln P_{\mathrm{FA}}).
\label{eq:sensing_quality_core}
\end{align}
Here $P_{\mathrm{FA}}$ is false-alarm probability and $F_{\chi'^2_2(\zeta)}$ the noncentral-$\chi^2$ CDF with two degrees of freedom and noncentrality $\zeta$. Localization uses RMS bandwidth $\beta_p=B_p^{\mathrm s}/\sqrt{12}$ and variances $\sigma_d^2=c_0^2/(32\pi^2\beta_p^2\gamma_{j,p,t}^{\mathrm s})$ and $\sigma_\phi^2=(2\gamma_{j,p,t}^{\mathrm s}\kappa_\phi)^{-1}$, where $c_0$ is propagation speed, $D_{\mathrm{eff}}$ effective aperture, and $\kappa_\phi=\frac{1}{12}(2\pi D_{\mathrm{eff}}/\lambda_{\mathrm c})^2$ is the bearing Fisher coefficient. Let $\mathbf H_{j,t}^{\mathrm{pol}}$, $\mathbf F$, $\mathbf Q$, and $\mathbf H^{\mathrm{trk}}$ denote the polar-position Jacobian, tracking transition, process covariance, and position selector; $\mathbf J_{j,p,t}^{\mathrm{pos}}$, $\mathbf P^-_{j,t}$, and $\mathbf P^+_{j,t}$ are the position information matrix and predicted/posterior covariances. The resulting position error bound (PEB) and session-level posterior Cram\'er--Rao bound (PCRB) are
\begin{align}
\mathbf J_{j,p,t}^{\mathrm{pos}}
&=(\mathbf H_{j,t}^{\mathrm{pol}})^\top
\operatorname{diag}(\sigma_d^2,\sigma_\phi^2)^{-1}\mathbf H_{j,t}^{\mathrm{pol}},\nonumber\\
\operatorname{PEB}_{j}(p,t)
&=\sqrt{\operatorname{tr}[(\mathbf J_{j,p,t}^{\mathrm{pos}})^{-1}]},\qquad
\mathbf P^-_{j,t}=\mathbf F\mathbf P^+_{j,t-1}\mathbf F^\top+\mathbf Q,\nonumber\\
\mathbf P^+_{j,t}
&=\left[(\mathbf P^-_{j,t})^{-1}+(\mathbf H^{\mathrm{trk}})^\top
\mathbf J_{j,p,t}^{\mathrm{pos}}\mathbf H^{\mathrm{trk}}\right]^{-1},\nonumber\\
\operatorname{PCRB}_{j}(t)
&=\sqrt{\operatorname{tr}\!\left(
\mathbf H^{\mathrm{trk}}\mathbf P^+_{j,t}
(\mathbf H^{\mathrm{trk}})^\top
\right)}.
\label{eq:localization_tracking_quality}
\end{align}
When no valid measurement update exists, the tracking posterior retains the current prior covariance. Detection requires $P_{\mathrm D,j}\ge\theta_r$; localization and tracking additionally require the detection gate and, respectively, $\operatorname{PEB}_{j}(p,t)\le\theta_r$ or $\operatorname{PCRB}_{j}(t)\le\theta_r$. Target-indexed primitive channel terms are shared by simultaneous sessions on the same physical target before profile allocation.

\subsection{Compatibility, Actions, and Service Accounting}

A feasible merge must preserve every member: exact target identity, focal-AOI coverage $\omega_{rj}=|A_r\cap A_j|/|A_r|\ge\omega_{\min}$, where $\omega_{\min}$ is the minimum coverage ratio, the target inside both AOIs, task-output capability $\kappa_r\in\mathcal O_j$, all member sharing permissions and cross-tenant authorization, valid start/lifetime, and freshness $\nu_p-1\le\Delta_{r'}$ for all $r'\in\mathcal S_{j,t}\cup\{r\}$. The candidate shared output must satisfy every member's current quality requirement and deterministic reservations. If $\mathcal U_j$ is session $j$'s update calendar, each reserved slot $\tau$ satisfies
\begin{equation}
\sum_{j:\tau\in\mathcal U_j}B_{p_{j,t}}^{\mathrm s}\le B^{\mathrm{tot}},\qquad
\sum_{j:\tau\in\mathcal U_j}P_{p_{j,t}}^{\mathrm s}\le P^{\mathrm{tot}}.
\label{eq:future_reservation}
\end{equation}
Unknown future arrivals are not reserved. Let $F_{rjp,t}$ denote the full merge conjunction; $F_{rp,t}^{\mathrm{create}}$ analogously checks start/lifetime, target/AOI and current-quality conditions, freshness, and reservation feasibility.

The feasible action set is
\begin{equation}
\begin{aligned}
\mathcal A_t={}&\{\textsc{Merge}(j,p):F_{rjp,t}=1\}\\
&\cup\{\textsc{Create}(p):F_{rp,t}^{\mathrm{create}}=1\}\\
&\cup\{\textsc{Defer}:t+\tau_{\mathrm{defer}}\le d_r\}\cup\{\textsc{Reject}\}.
\end{aligned}
\label{eq:action_space}
\end{equation}
\textsc{Merge}/\textsc{Create} admit the focal request into an existing or new session under $p$; \textsc{Defer} advances its next eligibility by the fixed cooldown $\tau_{\mathrm{defer}}$, and \textsc{Reject} terminates it. Hard masks encode only action feasibility, never learned value, reward, cost, or heuristic preference.

Admission starts with no fictitious valid result before the mandatory current-slot update. Let $A_{r,t}^{\mathrm{age}}$ and $N_{r,t}^{\mathrm{valid}}$ denote output age and accumulated valid outputs; a valid result resets age to zero, otherwise age increments. With $U_{r,t}^{\mathrm{SLA}}$ the absorbing post-admission violation flag, $\mathbb I[\cdot]$ the indicator, $B_{r,t}^{\mathrm{first}}$ the first-violation indicator, and $Z_r^{\mathrm{succ}}$ terminal success:
\begin{align}
B_{r,t}^{\mathrm{first}}
&=\mathbb I\!\left[U_{r,t}^{\mathrm{SLA}}=0\ \land\left(A_{r,t+1}^{\mathrm{age}}>\Delta_r\right.\right.\nonumber\\[-1mm]
&\hspace{26mm}\left.\left.\lor\ (t=f_r\land N_{r,t+1}^{\mathrm{valid}}=0)\right)\right],\nonumber\\
Z_r^{\mathrm{succ}}
&=\mathbb I\!\left[N_{r,f_r+1}^{\mathrm{valid}}\ge1\ \land\ U_{r,f_r+1}^{\mathrm{SLA}}=0\right].
\label{eq:service_accounting}
\end{align}
A first violation is absorbing but the request remains attached through $f_r$; $Z_r^{\mathrm{succ}}$ determines \textsc{Completed} versus \textsc{Failed}, while rejected/expired requests receive no completion value.

\subsection{Reward, Constraints, and CMDP}

Let $V_t^{\mathrm{comp}}=\sum_{r:f_r=t}v_rZ_r^{\mathrm{succ}}$. With bandwidth/power cost weights $w_B,w_P$ and sensing-cost weight $\lambda_{\mathrm{res}}$, the normalized sensing cost and slot reward are
\begin{align}
C_t^{\mathrm{sense}}
&=w_B\frac{B_t^{\mathrm s}}{B^{\mathrm{tot}}}+w_P\frac{P_t^{\mathrm s}}{P^{\mathrm{tot}}},\qquad w_B+w_P=1,\nonumber\\
R_t&=V_t^{\mathrm{comp}}-\lambda_{\mathrm{res}}C_t^{\mathrm{sense}}.
\label{eq:slot_reward}
\end{align}
There is no merge, acceptance, deferral, fairness, or compatibility bonus; consolidation changes return only through completed service and sensing expenditure.

For tenant $l$ with violation-rate budget $\delta_l$ and user $k$ with minimum rate $R_k^{\min}$ and shortfall budget $\epsilon_k$, let $I_{r,t}^{\mathrm{acc}}$ indicate current-slot admission; the additive residuals are
\begin{align}
c_{l,t}^{\mathrm{SLA}}
&=\sum_{r:l_r=l}B_{r,t}^{\mathrm{first}}-\delta_l\sum_{r:l_r=l}I_{r,t}^{\mathrm{acc}},\nonumber\\
\widetilde R_{k,t}^{\min}&=\min\{D_{k,t},R_k^{\min}\},\nonumber\\
S_{k,t}^{\mathrm c}
&=
\begin{cases}
\dfrac{[\widetilde R_{k,t}^{\min}-R_{k,t}^{\mathrm c}]_+}
{\widetilde R_{k,t}^{\min}}, & D_{k,t}>0,\\
0, & D_{k,t}=0,
\end{cases}\nonumber\\
c_{k,t}^{\mathrm{comm}}
&=\mathbb I[D_{k,t}>0]\bigl(S_{k,t}^{\mathrm c}-\epsilon_k\bigr),
\label{eq:cmdp_residuals}
\end{align}
where $[x]_+=\max\{x,0\}$, $\widetilde R_{k,t}^{\min}$ is the demand-capped target, and $S_{k,t}^{\mathrm c}$ the normalized communication shortfall; both $S_{k,t}^{\mathrm c}$ and $c_{k,t}^{\mathrm{comm}}$ are zero when $D_{k,t}=0$. With tenant-$l$ episode counts $N_l^{\mathrm{violated}}$ and $N_l^{\mathrm{accepted}}$, $\sum_t c_{l,t}^{\mathrm{SLA}}=N_l^{\mathrm{violated}}-\delta_lN_l^{\mathrm{accepted}}$. Latest-start deadlines are hard pre-admission conditions; communication shortfall is controlled by the long-term residual, not a hard mask.

Let $s_t$ denote the Markov state, containing current physical processes, request/lifecycle state, active sessions, deterministic commitments, and cumulative service/constraint accounting. The policy observes a normalized public transformation $o_t=\psi(s_t)$ containing current request, session, and resource features, current compatibility/quality margins, accounting, and hard masks; future arrivals and other future exogenous realizations are excluded, and identifiers serve only as relational keys. CT-PPO and JC-PPO share this observation and the action semantics above.

The finite-horizon control objective is
\begin{subequations}
\label{eq:cmdp}
\begin{align}
\max_{\pi}\quad &J_R(\pi)=\mathbb E_\pi\!\left[\sum_{t=0}^{T-1}R_t\right],\\
\mathrm{s.t.}\quad &\mathbb E_\pi\!\left[\sum_{t=0}^{T-1}c_{l,t}^{\mathrm{SLA}}\right]\le0,\quad \forall l\in\mathcal L,\\
&\mathbb E_\pi\!\left[\sum_{t=0}^{T-1}c_{k,t}^{\mathrm{comm}}\right]\le0,\quad \forall k\in\mathcal K,
\end{align}
\end{subequations}
subject to the hard feasibility masks. This keeps action validity, long-term service constraints, and the completed-value-versus-sensing-cost objective explicitly distinct.

\section{Common-Trace Factorized Constrained PPO}
\label{sec:method}

\subsection{Shared Policy Structure and Matched Reference}

\begin{figure}[h]
\centering
\includegraphics[width=\columnwidth]{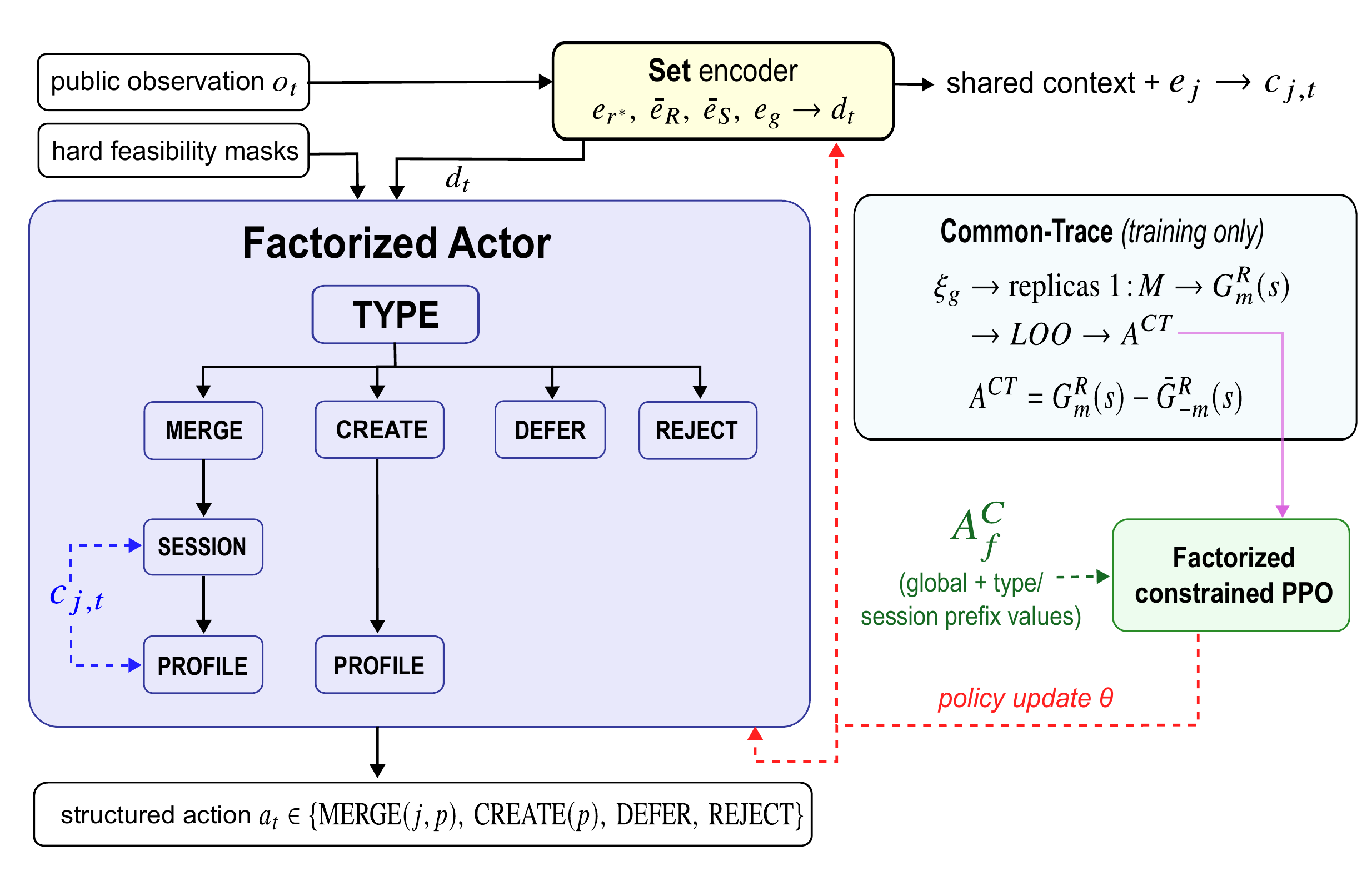}
\caption{CT-PPO architecture and training-time credit flow. The Set encoder and hard masks feed the factorized actor; leave-one-out common-trace reward credit $A^{\mathrm{CT}}$ and factor-specific constraint credit $A_f^C$ drive the constrained PPO update. Replication is used only during training. Schematic labels suppress indices: $(M,G_m^R,\bar G_{-m}^R,A_f^C)$ correspond to $(M_g,G_{g,m}^R,B_g^{R,-m},A_{n,f}^{C_q})$ below.}
\label{fig:ct_ppo_architecture}
\end{figure}

CT-PPO and JC-PPO use the same public observation, hard feasibility masks, Set representation, and conditional action distribution. Separate multilayer perceptrons (MLPs) embed request, session, and global features as $\mathbf e_r$, $\mathbf e_j$, and $\mathbf e_g$. With waiting-request set $\mathcal R_t$ and focal request $r^\star$, masked-mean summaries yield
\begin{equation}
\begin{aligned}
\bar{\mathbf e}_R&=\operatorname{mean}_{r\in\mathcal R_t}\mathbf e_r,\qquad
\bar{\mathbf e}_S=\operatorname{mean}_{j\in\mathcal J_t}\mathbf e_j,\\
\mathbf d_t&=\phi_D([\mathbf e_{r^\star},\bar{\mathbf e}_R,\bar{\mathbf e}_S,\mathbf e_g]),\\
\mathbf c_{j,t}&=\phi_M([\mathbf e_{r^\star},\mathbf e_j,\bar{\mathbf e}_R,\bar{\mathbf e}_S,\mathbf e_g]),
\end{aligned}
\label{eq:set_decision_embeddings}
\end{equation}
where $\phi_D$ and $\phi_M$ are learned MLP maps and the empty-session summary is zero. $\mathbf d_t$ supplies shared decision/critic context and $\mathbf c_{j,t}$ conditions the \textsc{Merge} session/profile branches; Fig.~\ref{fig:ct_ppo_architecture} summarizes the architecture.

The masked actor follows the environment hierarchy,
\begin{align}
\pi_\theta(a_t|o_t)
&=\pi_{T,\theta}(\tau_t|o_t)
\pi_{S,\theta}(j_t|o_t,\tau_t)^{I_{S,t}}\nonumber\\
&\quad\times\pi_{P,\theta}(p_t|o_t,\tau_t,j_t)^{I_{P,t}},
\label{eq:factorized_policy}
\end{align}
where $\theta$ denotes policy parameters, $\tau_t\in\{\textsc{Merge},\textsc{Create},\textsc{Defer},\textsc{Reject}\}$ the action type, and $I_{S,t},I_{P,t}$ session/profile applicability. Type always applies; session applies only to \textsc{Merge} with multiple feasible destinations, and profile only to \textsc{Merge}/\textsc{Create} with multiple feasible profiles; $j_t$ is omitted under \textsc{Create}. Singleton branches are deterministic, have zero conditional log probability, and receive no actor credit.

All factors are normalized after hard masking. For feasible \textsc{Merge} pair set $\mathcal F_t^M=\{(j,p):F_{rjp,t}=1\}$ and session/profile logits $u_S,u_{MP}$, the effective type logit includes
\begin{equation}
\Delta_t^{M}=\log\!\left[
\frac{1}{|\mathcal F_t^{M}|}
\sum_{(j,p)\in\mathcal F_t^{M}}
\exp\{u_S(j)+u_{MP}(j,p)\}
\right],
\label{eq:merge_option_score}
\end{equation}
which removes a feasible-pair multiplicity bonus. For decision $n$, $f\in\{T,S,P\}$ denotes type/session/profile, with selected component $a_{n,f}$ and applicability $I_{n,f}$. Let $\ell_{n,f}(\theta)=\log\pi_{f,\theta}(a_{n,f}|o_n)$ and $\ell_n^{\mathrm{joint}}(\theta)=\sum_f I_{n,f}\ell_{n,f}(\theta)$. The global critic predicts $V_R(o_n)$ and $Q=L+K$ values $V_{C_q}(o_n)$.

JC-PPO is the matched learned reference. A focal transition may span $h_n$ physical slots, so the shared rollout backbone uses span-aware GAE~\cite{schulman2016gae} with discount $\gamma$, parameter $\lambda$, bootstrap $\gamma^{h_n}$, and recursion $(\gamma\lambda)^{h_n}$ for reward and constraints. JC-PPO uses these global advantages with a clipped joint-ratio primal--dual actor. CT-PPO retains the same global critics and rollout/dual infrastructure but changes the actor-side credit and surrogate construction described below. Otherwise the methods share environment, observations, hard masks, Set encoder, action distribution, reward, constraints, optimizer family, training budget, validation protocol, and evaluation workloads.

\subsection{Common-Trace Reward Credit}

CT-PPO conditions reward credit on a shared exogenous workload realization. We call the policy-independent sequence of request arrivals and physical-process realizations a \emph{primitive workload trace}. A training group $g$ contains $M_g\ge2$ replicas sharing the same trace $\xi_g$; replicas draw their own policy actions, so endogenous request/session trajectories can diverge. In the final runs, each $5000$-slot rollout contains $25$ episodes, grouped as one triplet and eleven pairs. For replica $m$, define the discounted slot-return suffix, leave-one-out peer baseline, and decision credit as
\begin{align}
G^R_{g,m}(s)
&=\sum_{u=s}^{T-1}\gamma^{u-s}R_{g,m,u},\nonumber\\
B_g^{R,-m}(s)
&=\frac{1}{M_g-1}\sum_{m'\ne m}G^R_{g,m'}(s),\nonumber\\
A^{\mathrm{CT}}_{g,m,n}
&=G^R_{g,m}(s_{g,m,n})-B_g^{R,-m}(s_{g,m,n}),
\label{eq:ct_advantage}
\end{align}
where $R_{g,m,u}$ is the slot reward in~\eqref{eq:slot_reward} for replica $m$ in group $g$, and $s_{g,m,n}$ the decision's physical start slot. Peers need not decide there: alignment is by physical slot in the shared trace, not by endogenous state/action sequence, and replica $m$ is excluded from its baseline. After rollout normalization, $\widehat A_n^{\mathrm{CT}}$ is supplied only to applicable action factors.

\subsection{Factor-Specific Constraint Credit}

Constraint credit uses detached hierarchical prefix values. Here, a prefix is the set of higher-level choices already fixed before a lower factor is selected. The prefix is the action type for session selection, and the action type plus the selected session (under \textsc{Merge}) for profile selection. For $x\in\{R,C_1,\ldots,C_Q\}$,
\begin{align}
V_x^T(o_n,\tau)
&=\operatorname{sg}[V_x(o_n)]+\Delta_x^T(\operatorname{sg}[\mathbf d_n],\tau),\nonumber\\
V_x^S(o_n,j)
&=\operatorname{sg}[V_x^T(o_n,\textsc{Merge})]
+\Delta_x^S(\operatorname{sg}[\mathbf c_{j,n}]),
\label{eq:prefix_values}
\end{align}
where $\operatorname{sg}[\cdot]$ stops gradients and $\Delta_x^T,\Delta_x^S$ are residual type/session-prefix heads with zero-initialized outputs; prefix losses update only these heads.

Let $G_n^{C_q}$ be the rollout-frozen GAE return and $A_n^{C_q}$ the global rollout advantage for constraint $q$. Type retains this global constraint advantage, whereas lower factors subtract the corresponding old-policy prefix value:
\begin{align}
A_{n,T}^{C_q}&=A_n^{C_q},\nonumber\\
A_{n,S}^{C_q}&=I_{n,S}\!
\left[G_n^{C_q}-V_{C_q}^{T,\mathrm{old}}(o_n,\textsc{Merge})\right],\nonumber\\
A_{n,P}^{C_q}&=I_{n,P}
\begin{cases}
G_n^{C_q}-V_{C_q}^{S,\mathrm{old}}(o_n,j_n),&\tau_n=\textsc{Merge},\\
G_n^{C_q}-V_{C_q}^{T,\mathrm{old}}(o_n,\textsc{Create}),&\tau_n=\textsc{Create}.
\end{cases}
\label{eq:factor_constraint_credit}
\end{align}
Inapplicable lower-factor advantages are zero. These rollout-frozen advantages use the same rollout centering and scaling statistics as the corresponding global advantages. Reward prefix heads are trained on reward returns, but CT-PPO's actor reward term uses only the normalized common-trace advantage.

\subsection{Factorized Constrained PPO Update}

For each applicable factor, let $\theta_{\mathrm{old}}$ denote the rollout-policy parameters and $\epsilon$ the PPO clip ratio; define $\rho_{n,f}=\exp(\ell_{n,f}(\theta)-\ell_{n,f}(\theta_{\mathrm{old}}))$ and $\bar\rho_{n,f}=\operatorname{clip}(\rho_{n,f},1-\epsilon,1+\epsilon)$. The factorized reward and constraint surrogates are
\begin{align}
\mathcal L_R^{\mathrm{CT}}
&=\frac{1}{N}\sum_{n,f}I_{n,f}
\min\!\left(\rho_{n,f}\widehat A_n^{\mathrm{CT}},\bar\rho_{n,f}\widehat A_n^{\mathrm{CT}}\right),\nonumber\\
\mathcal L_{C_q}
&=\frac{1}{N}\sum_{n,f}I_{n,f}
\max\!\left(\rho_{n,f}\widehat A_{n,f}^{C_q},\bar\rho_{n,f}\widehat A_{n,f}^{C_q}\right),
\label{eq:ct_surrogates}
\end{align}
The denominator is the full minibatch size $N$, so sparse lower factors contribute in proportion to applicability; hats denote rollout-normalized advantages. With rollout scales $s_R,s_{C_q}$, raw dual $\lambda_q\ge0$, and entropy coefficient $\beta_H$, the actor uses $\widetilde\lambda_q=\lambda_q s_{C_q}/s_R$ and minimizes
\begin{equation}
\mathcal J_{\mathrm{actor}}
=-\mathcal L_R^{\mathrm{CT}}
+\sum_{q=1}^{Q}\widetilde\lambda_q\mathcal L_{C_q}
-\beta_H\mathcal H(\pi_\theta).
\label{eq:ct_actor_objective}
\end{equation}
Here $\mathcal H(\pi_\theta)$ is the entropy of the exact masked hierarchical policy. Factor-specific ratios drive the actor update; the reconstructed joint ratio is retained only for joint diagnostics, including approximate Kullback--Leibler (KL) divergence and clipping statistics, and for KL-based PPO early stopping.

Global and prefix critics use clipped value losses against rollout returns. Actor and global-critic gradients update the shared encoder with their respective heads; detached prefix losses update only prefix heads. After each rollout, both methods project raw duals using the mean episode residual: $\lambda_q\leftarrow\Pi_{[0,\lambda_{\max}]}[\lambda_q+\eta_\lambda E^{-1}\sum_{e=1}^{E}C_{e,q}]$, where $e=1,\ldots,E$ indexes rollout episodes, $C_{e,q}$ is the episode-total residual, $\eta_\lambda$ the dual learning rate, $\lambda_{\max}$ the cap, and $\Pi_{[0,\lambda_{\max}]}$ denotes projection onto $[0,\lambda_{\max}]$.

Common-trace returns and prefix values are training-only quantities. Validation, external evaluation, and deployment use a single trajectory with only the current public observation, hard masks, trained Set encoder, and factorized actor; peer trajectories/returns, prefix critics, future arrivals, regime labels, oracle outcomes, and hindsight information are absent from action selection.

\section{Experimental Methodology}
\label{sec:experiments}

\subsection{Experimental Setting and Online References}
\label{subsec:simulation_setting}

The primary comparison and component ablation are evaluated in the frozen nominal environment of Section~\ref{sec:system_model}; the arrival-load stress evaluation changes only the two arrival-rate scalars specified below. The four learned methods share the same public observations, hard masks, action semantics, Set encoder, conditional factorized action distribution, PPO/CMDP hyperparameters, validation protocol, and physical interaction budget; their training differences are defined in Section~\ref{subsec:extended_evaluation}. A workload root is a seed that, together with the arrival regime, deterministically generates one primitive workload trace; training seeds separately index learned-policy runs. Table~\ref{tab:experimental_setup} reports the nominal workload, model, optimization, and evaluation settings; learned-run entries are taken from the effective configurations stored with the final artifacts.

\begin{table*}[t]
\caption{Final experimental configuration. Learned-run values are taken from the effective configurations recorded in the final artifacts; $U[a,b]$ denotes the continuous uniform distribution on $[a,b]$.}
\label{tab:experimental_setup}
\centering
\footnotesize
\renewcommand{\arraystretch}{1.06}
\begin{tabular}{@{}p{0.18\textwidth}p{0.76\textwidth}@{}}
\toprule
\textbf{Category} & \textbf{Setting} \\
\midrule
System & $T=200$ slots, $\Delta t=0.1$ s; $L=4$ sensing tenants, $K=6$ communication users, $8$ targets; $B^{\mathrm{tot}}=20$ MHz, $P^{\mathrm{tot}}=40$ W; carrier $6$ GHz. \\
Geometry/mobility & $[-200,200]^2$ m$^2$ region; initial target/user ranges $30$--$140$/$20$--$180$ m; initial target/user speeds $0$--$12$/$0$--$8$ m/s and acceleration std. $1/0.5$ m/s$^2$; AOI radius $U[15,30]$ m with center-offset std. $4$ m, truncated at half the AOI radius. \\
Requests & DET/LOC/TRK probabilities $0.35/0.35/0.30$; durations $H_{\mathrm{DET}}/H_{\mathrm{LOC}}/H_{\mathrm{TRK}}=3/4/8$ slots; completion values $U[0.8,1.2]/U[1.5,2.5]/U[2.5,4.0]$; latest-start slack discrete-uniform on $\{2,\ldots,8\}$; update periods DET/LOC on $\{1,2,3\}$ with probabilities $(.25,.45,.30)/(.30,.45,.25)$ and TRK on $\{1,2\}$ with $(.65,.35)$; sharing permission $0.9$; defer cooldown $\tau_{\mathrm{defer}}=1$ slot. \\
Physical/quality & $N_0=-174$ dBm/Hz; communication/sensing noise figures $7/7$ dB; communication gap $1.5$ dB; sensing front-end gain/system loss $24/3$ dB, effective aperture $0.5$ m; $P_{\mathrm{FA}}=10^{-4}$, detection gate $0.9$; DET/PEB/PCRB thresholds $U[0.85,0.98]$/$U[1.5,6]$/$U[1.5,5]$ m; tracking prior covariance diagonal $(25,25,4,4)$. \\
Channel processes & Communication path-loss exponent $3$, shadowing std./correlation $4$ dB/$0.95$, fading correlation $0.9$; sensing RCS median $1$ m$^2$, RCS std./correlation $3$ dB/$0.98$, shadowing std./correlation $3$ dB/$0.95$, fading correlation $0.9$. \\
Communication & Minimum rate $R_k^{\min}=2$ Mbit/s; Markov on/off demand with initial-on $0.50$, on$\to$off/off$\to$on probabilities $0.08/0.20$, and lognormal on-demand median $5$ Mbit/s (log-std.\ $0.45$); equal-share scheduling. \\
Sharing/profiles & Minimum request-to-session spatial coverage ratio $\omega_{\min}=0.80$; cross-tenant sharing is disallowed only for tenant pairs $(1,4)$ and $(2,3)$ (and symmetric counterparts). Profiles: economical $(2\,\mathrm{MHz},2\,\mathrm{W},3)$; balanced $(4\,\mathrm{MHz},5\,\mathrm{W},2)$; precision $(8\,\mathrm{MHz},8\,\mathrm{W},2)$; rapid $(4\,\mathrm{MHz},8\,\mathrm{W},1)$, where the third entry is the update period. \\
Objective/constraints & $\lambda_{\mathrm{res}}=0.2$, bandwidth/power cost weights $w_B=w_P=0.5$, $\gamma=1$; per-tenant sensing-SLA budget $\delta_l=0.05$ and per-user communication-shortfall budget $\epsilon_k=0.05$. \\
Arrival regimes & Independent: per-tenant Poisson rate $0.08$/slot. Clustered: parent Poisson rate $0.08$/slot, $1+\mathrm{Poisson}(2)$ children, discrete-uniform offsets $\{0,1,2,3\}$, parent-target/task inheritance $0.9/0.6$. \\
Training/model & Seeds $0$--$4$ per method; $10^6$ physical slots/run; rollout $5000$ slots; Adam ($\epsilon_{\mathrm{Adam}}=10^{-5}$), initial LR $3\times10^{-4}$ with linear $10^6$-slot schedule; $10$ epochs, minibatch $512$; hidden/profile dimensions $128/32$, $\tanh$, masked-mean pooling, orthogonal initialization, zero dropout; first-rollout feature normalization then frozen (clip $10$, $\epsilon_{\mathrm{norm}}=10^{-8}$); deterministic CPU, one Torch thread. \\
PPO/CMDP & PPO clip $\epsilon=0.2$, value clip $0.2$, GAE $\lambda=0.95$, entropy coefficient $\beta_H=0.01$, reward/constraint value coefficients $0.5/0.5$, target KL $0.03$, max gradient norm $0.5$; dual LR $\eta_\lambda=0.01$, maximum $\lambda_{\max}=100$. \\
Validation/evaluation & Validation roots $51001$--$51020$ in both regimes every $10^4$ slots; Random Valid root $53001$, four replicates/trace; external roots $52001$--$52050$ in both regimes; bootstrap root $54001$, $10^4$ samples. \\
\bottomrule
\end{tabular}
\end{table*}

Independent arrivals use a separate Poisson process for each tenant. Clustered arrivals use common parent events with correlated temporal offsets and inherited target/task attributes, creating more session-reuse opportunities without exposing the regime label to the learned policy.

The reference set comprises JC-PPO, four deterministic online heuristics, and Random Valid. \emph{No Consolidation} disables \textsc{Merge} and selects the feasible \textsc{Create} using the lowest-cost sensing profile. \emph{Static Compatibility Merge} prioritizes \textsc{Merge} and selects the feasible merge using the lowest-cost sensing profile, falling back to the corresponding lowest-cost \textsc{Create}. \emph{Greedy Incremental Cost} minimizes the increase in current normalized sensing-resource cost over feasible immediate-service actions. \emph{SLA-Aware Greedy} first maximizes the worst current normalized quality/freshness margin among the affected requests and then minimizes incremental sensing-resource cost. If no immediate-service action exists, the heuristics defer when feasible and reject otherwise. \emph{Random Valid} samples uniformly from feasible actions and is used only as a stochastic sanity reference, not as a lower bound. JC-PPO is the matched learned reference described in Section~\ref{sec:method}.

\subsection{Training, Model Selection, and External Evaluation}
\label{subsec:training_protocol}

Five policies are trained for each learned method using training seeds $\{0,1,2,3,4\}$, each receiving exactly $10^6$ physical interaction slots. Every replica trajectory used by a common-trace method contributes to this budget, so common-trace grouping receives no additional physical interaction. All four learned methods use the same two regimes and Table~\ref{tab:experimental_setup} settings; all twenty learned runs completed the budget.

Checkpoint selection is isolated from external evaluation. At slot $0$ and every $10^4$ slots thereafter, the current deterministic policy is evaluated on roots $51001$--$51020$ in both regimes. The selection score is policy return minus a separately seeded Random Valid return on the same validation workloads. Only the top-ranked checkpoint is retained; checkpoints are ordered by macro paired-return difference, worst-regime paired-return difference, lower macro positive constraint excess, and earlier physical slot. The \emph{best} checkpoint for each training seed is therefore fixed without consulting external roots $52001$--$52050$; the final (\emph{latest}) checkpoint is retained only for the secondary stability analysis.

After selection, each best checkpoint is evaluated deterministically on roots $52001$--$52050$ in both regimes ($100$ episodes/policy). Heuristics run once on the same root--regime traces, while Random Valid uses four independently seeded action replicates per trace. For each root, macro observations are formed by equally averaging the two regimes; the underlying records remain separated by training seed, root, regime, and stochastic replicate as applicable.

\subsection{Component Ablation, Arrival-Load Stress, and Deployment Evaluation}
\label{subsec:extended_evaluation}

To separate CT-PPO's policy-update components, we evaluate two intermediate variants under the same training, validation, checkpoint-selection, and external-evaluation protocol. \emph{Factorized-JC} retains JC-PPO's model and global reward/constraint GAE but replaces its clipped joint-ratio actor surrogate with the factor-wise surrogate used by CT-PPO; it uses neither common-trace reward credit nor prefix critics. \emph{CT-Reward} retains the Factorized-JC model and factor-wise surrogate, replaces its reward advantage with CT-PPO's leave-one-out Monte Carlo reward credit from the same primitive workload trace, aligned by physical slot, and retains global JC-style constraint GAE; it likewise has no prefix critics. Hence the adjacent contrasts along JC-PPO $\rightarrow$ Factorized-JC $\rightarrow$ CT-Reward $\rightarrow$ CT-PPO isolate, respectively, the factor-wise actor surrogate, common-trace reward credit after controlling the surrogate, and the prefix-critic machinery for factor-specific constraint credit added by CT-PPO.

To test arrival-load robustness without retraining or checkpoint reselection, the five frozen validation-selected CT-PPO/JC-PPO checkpoint pairs are evaluated at low and high arrival rates in addition to the nominal setting. The low/nominal/high configurations set the independent per-tenant Poisson rate and clustered parent-event rate to $0.06/0.08/0.10$ per slot, respectively; the low and high configurations differ from the nominal environment only in these two scalars. Each setting uses roots $52001$--$52050$ in both arrival regimes. Root identifiers are intentionally reused across load settings to preserve matched-root coupling; this is therefore a matched-root load-shift stress test rather than an independent fresh-root replication.

Computational and deployment evaluation separates trainable model size from action-selection latency. Trainable parameters are counted directly from the instantiated models and partitioned into the Set encoder, policy head, global critic, and prefix critic. Inference is measured for the current full action-selection implementation (the reference path) and a deployment-equivalent actor-only path consisting of feature normalization, Set encoding, policy-head evaluation, and deterministic action selection. The same $500$ public observations, generated by running SLA-Aware Greedy in the nominal environment and spanning both arrival regimes, are presented to CT-PPO and JC-PPO for all five frozen checkpoint pairs. Before timing, actor-only actions are checked against the reference path on the first $100$ observations for each checkpoint pair. Measurements use CPU execution, one Torch thread, batch size one, $100$ warm-up observations, and alternating JC-PPO/CT-PPO timing order across successive observations.

\subsection{Metrics and Statistical Analysis}
\label{subsec:statistics}

Episode return under~\eqref{eq:slot_reward} is the primary endpoint. Its service--resource decomposition uses completed value and sensing-resource cost. Consolidation is measured directly by \textsc{Merge}/\textsc{Create} counts and $\mathrm{RPS}=N^{\mathrm{accepted}}/N^{\mathrm{create}}$ when $N^{\mathrm{create}}>0$, where RPS means \emph{accepted requests per created physical sensing session}. Constraint performance is reported by family-specific episode residuals and their positive excess
\begin{equation}
E^+
=
\sum_{l\in\mathcal L}\left[\sum_t c_{l,t}^{\mathrm{SLA}}\right]_+
+
\sum_{k\in\mathcal K}\left[\sum_t c_{k,t}^{\mathrm{comm}}\right]_+.
\label{eq:positive_constraint_excess}
\end{equation}
Positive $E^+$ is reported as achieved constraint excess, not interpreted as strict feasibility.

The five training seeds are the independent learned-policy units; the $50$ external roots and two regimes are repeated matched workloads within each trained policy. All learned-to-learned paired effects reported below---the primary CT-PPO--JC-PPO comparison, the three adjacent component-ablation contrasts, and the CT-PPO--JC-PPO comparison at each arrival-load setting---use the same $10^4$-draw hierarchical paired bootstrap. Each draw resamples five matched training-seed indices with replacement and, within each sampled seed pair, resamples the $50$ matched roots; the two regime outcomes for a root remain paired and are equally weighted for macro inference. The reported statistic is the mean paired effect across the resampled seed blocks with percentile $95\%$ confidence intervals. Component attribution uses only adjacent contrasts along JC-PPO $\rightarrow$ Factorized-JC $\rightarrow$ CT-Reward $\rightarrow$ CT-PPO, avoiding nonadjacent differences that change multiple training components. CT-PPO--heuristic effects use the same hierarchy for the CT training seed and roots while pairing each CT outcome with the deterministic heuristic on the identical primitive trace.

For absolute operating-point intervals in Fig.~\ref{fig:consolidation_mechanism}, learned policies are bootstrapped over training seeds then roots, while heuristic intervals resample roots only; comparisons use paired effects rather than error-bar overlap. Random Valid is averaged over four action replicates within each trace. The evaluation protocol fixes bootstrap root $54001$ and $10^4$ samples. Fig.~\ref{fig:validation_dynamics}'s min--max seed bands are descriptive, not confidence intervals, and repeated roots are never treated as independent training experiments. 

\section{Results}
\label{sec:results}

\subsection{External Performance and Practical References}
\label{subsec:overall_results}

Table~\ref{tab:overall_external_results} reports the macro external operating points. CT-PPO attains the highest mean return, $67.511$, ahead of JC-PPO ($66.576$) and SLA-Aware Greedy ($65.663$). Static Compatibility Merge and Greedy Incremental Cost reduce sensing cost to about $16.3$, but completed value falls to about $66.7$ and positive constraint excess exceeds $4.6$, showing that low sensing expenditure alone does not define an effective consolidation policy.

\begin{table*}[t]
\caption{Primary external macro results. Learned entries average five validation-selected training seeds; deterministic heuristics average matched external roots, and Random Valid is averaged within trace. $E^+$ is positive constraint excess; RPS is accepted requests per created session.}
\label{tab:overall_external_results}
\centering
\footnotesize
\renewcommand{\arraystretch}{1.05}
\begin{tabular}{@{}lrrrrr@{}}
\toprule
\textbf{Policy} & \textbf{Return} & \textbf{Completed value} & \textbf{Sense cost} & \boldmath$\mathbf{E^+}$ & \textbf{RPS} \\
\midrule
CT-PPO & \textbf{67.511} & 74.005 & 32.474 & 2.164 & 1.114 \\
JC-PPO & 66.576 & 74.326 & 38.751 & 2.382 & 1.034 \\
SLA-Aware Greedy & 65.663 & 72.678 & 35.076 & 2.512 & 1.081 \\
Static Compatibility Merge & 63.438 & 66.702 & 16.321 & 4.627 & 1.125 \\
Greedy Incremental Cost & 63.417 & 66.673 & 16.278 & 4.641 & 1.105 \\
No Consolidation & 62.998 & 66.366 & 16.843 & 4.722 & 1.000 \\
Random Valid & 42.901 & 46.519 & 18.089 & 2.122 & 1.048 \\
\bottomrule
\end{tabular}
\end{table*}

Against SLA-Aware Greedy, the strongest deterministic reference by return, CT-PPO gains $1.847$ (95\% CI $[1.604,2.094]$). The macro effect combines $+1.327$ completed value $[0.995,1.657]$ with $-2.602$ sensing-resource cost $[-3.829,-1.198]$ and $-0.348$ positive constraint excess $[-0.491,-0.205]$; RPS also increases by $0.0332$ $[0.0231,0.0428]$. CT-PPO therefore improves both components of the slot reward in~\eqref{eq:slot_reward} relative to this practical reference while using shared sessions more effectively.

\subsection{Integrated CT-PPO--JC-PPO Effect and Consolidation Behavior}
\label{subsec:ct_vs_jc}

The matched CT--JC comparison in Table~\ref{tab:ct_jc_effects} evaluates the complete CT-PPO training design against JC-PPO while holding the environment, public observation, hard masks, Set representation, action distribution, reward and constraints, optimizer family, training budget, validation rule, and external traces fixed. Because the endpoints differ in both surrogate construction and actor credit, this remains an integrated-method comparison; Section~\ref{subsec:component_attribution} resolves the individual interventions. CT-PPO improves macro return by $0.934$ (95\% CI $[0.702,1.164]$), with a positive CT--JC macro-return difference for all five matched training-seed indices. The reward decomposition locates this gain primarily in resource efficiency: completed value changes by $-0.321$ $[-0.660,0.023]$, whereas sensing cost falls by $6.277$ $[-7.625,-4.804]$. With $\lambda_{\mathrm{res}}=0.2$, the mean effects satisfy $-0.321-0.2(-6.277)=+0.934$. Direct consolidation metrics move consistently with this decomposition: RPS rises by $0.0806$, \textsc{Merge} increases by $2.344$ actions/episode, and \textsc{Create} decreases by $2.272$.

\begin{table*}[t]
\caption{CT-PPO minus JC-PPO effects from validation-selected checkpoints. Entries are $\Delta$ [95\% hierarchical paired-bootstrap CI].}
\label{tab:ct_jc_effects}
\centering
\footnotesize
\renewcommand{\arraystretch}{1.05}
\begin{tabular}{@{}lccc@{}}
\toprule
\textbf{Metric} & \textbf{Macro} & \textbf{Independent} & \textbf{Clustered} \\
\midrule
Return & $+0.934\ [0.702,1.164]$ & $+0.827\ [0.573,1.072]$ & $+1.042\ [0.752,1.329]$ \\
Completed value & $-0.321\ [-0.660,0.023]$ & $-0.583\ [-1.037,-0.148]$ & $-0.059\ [-0.351,0.246]$ \\
Sensing-resource cost & $-6.277\ [-7.625,-4.804]$ & $-7.047\ [-8.984,-4.889]$ & $-5.507\ [-6.578,-4.399]$ \\
Positive constraint excess & $-0.219\ [-0.573,0.058]$ & $+0.227\ [0.088,0.409]$ & $-0.665\ [-1.371,-0.128]$ \\
RPS & $+0.0806\ [0.0656,0.0954]$ & $+0.0177\ [0.0134,0.0221]$ & $+0.1434\ [0.1164,0.1701]$ \\
\textsc{Merge}/episode & $+2.344\ [1.926,2.754]$ & $+0.892\ [0.684,1.104]$ & $+3.796\ [3.120,4.472]$ \\
\textsc{Create}/episode & $-2.272\ [-2.674,-1.860]$ & $-0.868\ [-1.072,-0.672]$ & $-3.676\ [-4.356,-2.988]$ \\
\bottomrule
\end{tabular}
\end{table*}

The regime split reveals where the consolidation mechanism is most effective. Under independent arrivals, CT-PPO improves return by $0.827$ while reducing sensing cost by $7.047$; completed value decreases by $0.583$, and positive constraint excess increases by $0.227$. Under clustered arrivals, return improves by $1.042$, sensing cost falls by $5.507$, the completed-value interval spans zero, and positive excess decreases by $0.665$. The corresponding RPS effect grows from $+0.0177$ under independent arrivals to $+0.1434$ under clustered arrivals, showing that CT-PPO makes greater use of the richer reuse opportunities created by clustered demand.

\begin{figure*}[t]
\centering
\includegraphics[width=\textwidth]{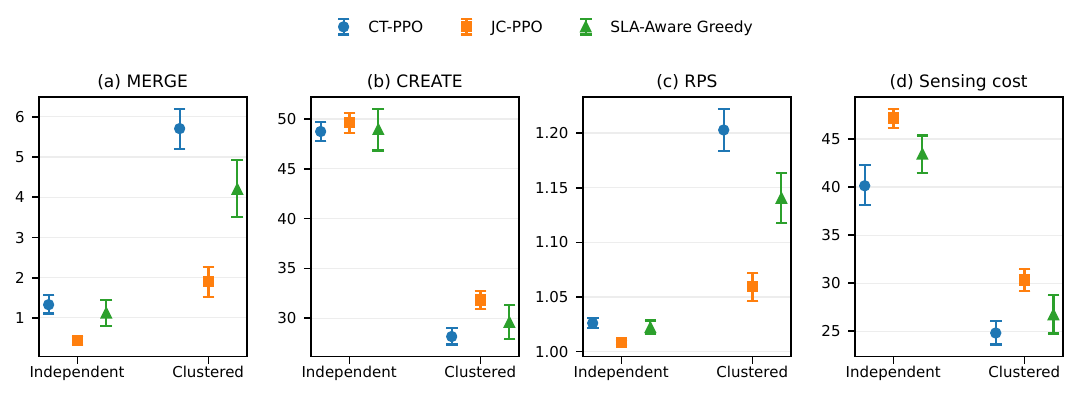}
\caption{Direct consolidation and sensing-resource behavior across arrival regimes. Points denote mean \textsc{Merge}/\textsc{Create} actions per episode, accepted requests per created session (RPS), and episode sensing-resource cost. Whiskers show percentile 95\% intervals from a hierarchical bootstrap over five training seeds and 50 external roots for learned policies, and a root-level bootstrap over 50 roots for SLA-Aware Greedy.}
\label{fig:consolidation_mechanism}
\end{figure*}

Figure~\ref{fig:consolidation_mechanism} complements these paired effects with the absolute consolidation operating points. The heuristic comparison also shows why reuse must be selective: Static Compatibility Merge attains slightly higher macro RPS than CT-PPO ($1.125$ versus $1.114$), but with substantially lower completed value ($66.702$ versus $74.005$) and larger positive constraint excess ($4.627$ versus $2.164$).

\subsection{Component Attribution}
\label{subsec:component_attribution}

The four-way ablation resolves the confounding present in the integrated CT--JC comparison by introducing one training change at each adjacent step. The external macro returns of the validation-selected checkpoints are $66.576$ for JC-PPO, $66.478$ for Factorized-JC, $67.490$ for CT-Reward, and $67.511$ for full CT-PPO. Table~\ref{tab:component_ablation} reports the corresponding adjacent paired effects.

\begin{table*}[t]
\caption{Adjacent component-ablation effects. Each entry is the second method minus the first method, $\Delta$ [95\% hierarchical paired-bootstrap CI]. Only adjacent contrasts are used for attribution.}
\label{tab:component_ablation}
\centering
\footnotesize
\renewcommand{\arraystretch}{1.05}
\begin{tabular}{@{}lccc@{}}
\toprule
\textbf{Metric} & \textbf{Factorized-JC $-$ JC-PPO} & \textbf{CT-Reward $-$ Factorized-JC} & \textbf{CT-PPO $-$ CT-Reward} \\
\midrule
Return & $-0.099\ [-0.472,0.287]$ & $+1.013\ [0.666,1.338]$ & $+0.021\ [-0.143,0.209]$ \\
Completed value & $-0.692\ [-1.272,-0.239]$ & $+0.343\ [-0.028,0.842]$ & $+0.028\ [-0.206,0.316]$ \\
Sensing-resource cost & $-2.968\ [-7.034,-0.064]$ & $-3.349\ [-6.233,0.265]$ & $+0.039\ [-0.585,0.693]$ \\
Positive constraint excess & $-0.016\ [-0.439,0.338]$ & $-0.186\ [-0.421,0.025]$ & $-0.016\ [-0.124,0.079]$ \\
RPS & $+0.0063\ [-0.0089,0.0250]$ & $+0.0625\ [0.0414,0.0818]$ & $+0.0117\ [-0.0031,0.0290]$ \\
\textsc{Merge}/episode & $+0.154\ [-0.344,0.752]$ & $+1.864\ [1.250,2.448]$ & $+0.326\ [-0.100,0.828]$ \\
\textsc{Create}/episode & $-0.276\ [-0.838,0.222]$ & $-1.658\ [-2.244,-0.968]$ & $-0.338\ [-0.852,0.096]$ \\
\bottomrule
\end{tabular}
\end{table*}

Replacing JC-PPO's joint-ratio surrogate with the factor-wise surrogate alone does not yield a detectable macro-return improvement: Factorized-JC minus JC-PPO is $-0.099$ $[-0.472,0.287]$, with intervals spanning zero in both independent ($-0.182$ $[-0.535,0.183]$) and clustered ($-0.015$ $[-0.523,0.496]$) arrivals. This intervention nevertheless changes the service--resource balance: completed value decreases by $0.692$, while sensing-resource cost decreases by $2.968$. Thus factorization alone does not explain the CT-PPO return gain.

After controlling for that surrogate change, common-trace reward credit produces the main return gain. CT-Reward improves macro return over Factorized-JC by $1.013$ $[0.666,1.338]$; the seedwise macro differences are positive for all five training seeds, and the regime-specific effects are $+0.994$ $[0.618,1.344]$ under independent arrivals and $+1.031$ $[0.605,1.490]$ under clustered arrivals. The isolated intervention also shifts behavior toward consolidation: RPS increases by $0.0625$ $[0.0414,0.0818]$, \textsc{Merge} by $1.864$ $[1.250,2.448]$, and \textsc{Create} decreases by $1.658$ $[-2.244,-0.968]$. Its mean sensing-cost effect is negative, but the corresponding interval $[-6.233,0.265]$ spans zero. These results identify common-trace reward credit, after controlling for the factorized surrogate, as the dominant empirical driver of the observed return improvement in this configuration.

Finally, adding CT-PPO's prefix-critic machinery for factor-specific constraint credit to CT-Reward changes macro return by only $+0.021$ $[-0.143,0.209]$. The intervals for completed value, sensing cost, positive constraint excess, RPS, and \textsc{Merge}/\textsc{Create} counts also span zero. The ablation therefore provides no detectable incremental return contribution from this additional machinery in the evaluated configuration; this is a non-detection, not an equivalence claim.

\subsection{Arrival-Load Robustness}
\label{subsec:load_robustness}

Table~\ref{tab:load_robustness} evaluates the same frozen validation-selected CT-PPO and JC-PPO checkpoints at the three tested arrival-rate settings. The CT--JC macro-return effect is positive at every load, increasing from $+0.690$ at rate $0.06$ to $+0.934$ at $0.08$ and $+1.443$ at $0.10$; all $15$ load-by-training-seed macro contrasts are positive.

\begin{table*}[t]
\caption{Frozen-checkpoint arrival-load robustness. The load is the common scalar applied to the independent per-tenant arrival rate and clustered parent-event rate. Operating-point rows report mean return; effect rows are CT-PPO minus JC-PPO, $\Delta$ [95\% hierarchical paired-bootstrap CI].}
\label{tab:load_robustness}
\centering
\footnotesize
\renewcommand{\arraystretch}{1.05}
\begin{tabular}{@{}lccc@{}}
\toprule
\textbf{Metric} & \textbf{Low ($0.06$)} & \textbf{Nominal ($0.08$)} & \textbf{High ($0.10$)} \\
\midrule
CT-PPO return & $52.631$ & $67.511$ & $83.978$ \\
JC-PPO return & $51.941$ & $66.576$ & $82.535$ \\
\midrule
Return & $+0.690\ [0.475,0.910]$ & $+0.934\ [0.702,1.164]$ & $+1.443\ [1.137,1.748]$ \\
Completed value & $-0.275\ [-0.486,-0.092]$ & $-0.321\ [-0.660,0.023]$ & $-0.061\ [-0.579,0.428]$ \\
Sensing-resource cost & $-4.828\ [-5.944,-3.613]$ & $-6.277\ [-7.625,-4.804]$ & $-7.521\ [-9.324,-5.685]$ \\
Positive constraint excess & $+0.075\ [0.006,0.160]$ & $-0.219\ [-0.573,0.058]$ & $-0.748\ [-1.134,-0.390]$ \\
RPS & $+0.0711\ [0.0541,0.0861]$ & $+0.0806\ [0.0656,0.0954]$ & $+0.0746\ [0.0633,0.0853]$ \\
\textsc{Merge}/episode & $+1.646\ [1.284,1.970]$ & $+2.344\ [1.926,2.754]$ & $+2.798\ [2.364,3.190]$ \\
\textsc{Create}/episode & $-1.630\ [-1.948,-1.278]$ & $-2.272\ [-2.674,-1.860]$ & $-2.608\ [-2.980,-2.196]$ \\
\bottomrule
\end{tabular}
\end{table*}

The return advantage also persists within each arrival regime at all three loads. At rate $0.06$, the independent and clustered effects are $+0.596$ $[0.361,0.805]$ and $+0.785$ $[0.472,1.066]$; at the nominal rate they are $+0.827$ $[0.573,1.072]$ and $+1.042$ $[0.752,1.329]$; and at rate $0.10$ they increase to $+1.166$ $[0.868,1.456]$ and $+1.719$ $[1.323,2.141]$. Across the three tested loads, the macro gain increases with load, with a larger increase under clustered traffic where reuse opportunities are richer; this trend is limited to the tested load grid and does not establish a universal monotonic law.

The decomposition also shows why the result should not be read as uniform dominance. At low load, CT-PPO reduces sensing-resource cost by $4.828$ and increases RPS by $0.0711$, but completed value decreases by $0.275$ and positive constraint excess increases by $0.075$, with both intervals excluding zero. The positive return effect at this load is therefore driven primarily by improved sensing-resource efficiency, with a modest service/constraint trade-off. At high load, completed value has no detectable change ($-0.061$ $[-0.579,0.428]$), while sensing cost decreases by $7.521$ and positive constraint excess by $0.748$; the consolidation shift remains consistent through more \textsc{Merge} and fewer \textsc{Create} actions. The frozen CT-PPO checkpoints thus retain their return advantage without retraining, while its source varies with load.

\subsection{Constraint and Training Diagnostics}
\label{subsec:tradeoffs}

Constraint behavior is family-specific. CT-PPO has macro $E^+=2.164$, entirely from sensing-SLA excess in the selected external episodes, whereas JC-PPO has $1.991$ sensing-SLA excess and $0.392$ communication-QoS excess. The lower mean total excess for CT-PPO therefore reflects zero observed communication-QoS excess together with a modestly higher sensing-SLA component, rather than uniform improvement across constraint families.

\begin{figure}[t]
\centering
\includegraphics[width=\columnwidth]{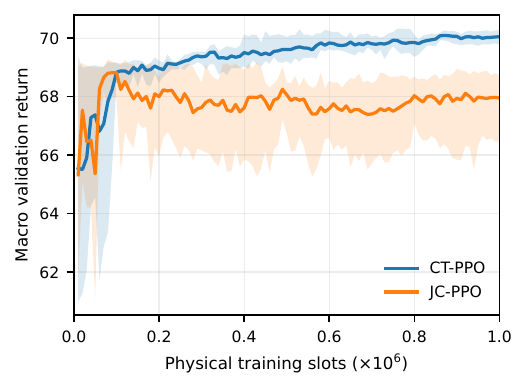}
\caption{Validation dynamics over physical training slots. Curves are unsmoothed five-seed mean macro validation returns; shading spans the seedwise minimum--maximum range.}
\label{fig:validation_dynamics}
\end{figure}

Figure~\ref{fig:validation_dynamics} shows the unsmoothed macro validation trajectory over the fixed $10^6$-slot training horizon. CT-PPO's selected checkpoints occur at $790$k--$1000$k slots; its mean latest-minus-best external return is $+0.037$ (latest $67.548$, selected $67.511$). JC-PPO is selected much earlier, at $20$k--$220$k slots, and all five runs end below their selected checkpoints, with mean latest-minus-best $-1.000$ (latest $65.577$, selected $66.576$). These dynamics support validation-only checkpoint selection; CT-PPO also retains its selected external performance at the endpoint on average, without implying asymptotic convergence.

\subsection{Deployment Footprint and Inference Cost}
\label{subsec:deployment_cost}

Table~\ref{tab:deployment_cost} separates training-time model capacity from the deployed action-selection path. JC-PPO, Factorized-JC, and CT-Reward each contain $306{,}450$ trainable parameters; full CT-PPO contains $346{,}569$, an increase of $40{,}119$ ($13.09\%$), exactly accounted for by its additional prefix-critic parameters. All four variants have the same $288{,}519$-parameter deployed encoder--actor footprint.

\begin{table}[t]
\caption{Model footprint and batch-one CPU inference. Latencies are five-checkpoint-pair means over the same $500$ environment-generated observations; actor-only actions were checked against the reference path before timing.}
\label{tab:deployment_cost}
\centering
\footnotesize
\renewcommand{\arraystretch}{1.04}
\begin{tabular}{@{}lrrr@{}}
\toprule
\multicolumn{4}{c}{\textbf{Trainable and deployed parameters}} \\
\cmidrule(lr){1-4}
\textbf{Method} & \textbf{Trainable} & \textbf{Prefix critic} & \textbf{Encoder--actor} \\
\midrule
JC-PPO & 306,450 & 0 & 288,519 \\
Factorized-JC & 306,450 & 0 & 288,519 \\
CT-Reward & 306,450 & 0 & 288,519 \\
CT-PPO & 346,569 & 40,119 & 288,519 \\
\midrule
\multicolumn{4}{c}{\textbf{Mean inference latency (ms)}} \\
\cmidrule(lr){1-4}
\textbf{Path} & \textbf{JC-PPO} & \textbf{CT-PPO} & \textbf{CT/JC} \\
\midrule
Reference path & 2.883 & 3.093 & 1.073 \\
Actor-only & 2.777 & 2.778 & 1.000 \\
\bottomrule
\end{tabular}
\end{table}

The reference path evaluates the actor and global critic and, for full CT-PPO, the prefix critic; its mean CT/JC latency ratio is $1.073$. Removing training-side critics gives the deployment-equivalent actor-only path---feature normalization, Set encoding, policy-head evaluation, and deterministic action selection---for which mean latency is $2.778$ ms for CT-PPO versus $2.777$ ms for JC-PPO, with a five-checkpoint-pair mean ratio of $1.0003$. Before timing, actor-only decisions matched the reference-path decisions on the first $100$ observations for both methods in every frozen checkpoint pair. Across the five pairs, actor-only CT/JC mean-latency ratios range from $0.9969$ to $1.0060$, whereas the reference-path ratios range from $1.0691$ to $1.0776$. Thus CT-PPO's additional trainable capacity is confined to training-side critic machinery: it increases measured latency on the full reference path but does not enlarge the deployed encoder--actor, and measured actor-only batch-one CPU latency is effectively unchanged. This evidence does not imply lower computational complexity or faster inference than JC-PPO.

\section{Discussion and Limitations}
\label{sec:discussion}

\subsection{Mechanism and Component Attribution}

The combined evidence separates the behavioral mechanism from the training intervention that produces it. At the policy level, CT-PPO shifts accepted demand toward reuse of existing physical sensing sessions and away from new session creation. Because the slot reward in~\eqref{eq:slot_reward} contains no explicit merge, reuse, or acceptance bonus, this behavior is beneficial only when sensing-resource savings compensate for any service loss. The heuristic operating points reinforce the same distinction: aggressive reuse or sensing-cost minimization can reduce resource expenditure while sacrificing completed value or constraint quality. The relevant control mechanism is therefore \emph{selective consolidation}, not consolidation for its own sake.

The four-way ablation sharpens the training-side attribution. Replacing JC-PPO's joint-ratio surrogate with the factor-wise surrogate alone changes the service--resource operating point but yields no detectable macro-return improvement. After that surrogate is held fixed, adding common-trace reward credit produces positive macro-return differences for all five matched training seeds and positive aggregate return effects in both arrival regimes, while increasing RPS and \textsc{Merge} and reducing \textsc{Create}. This connects the return gain to the common-trace reward-credit intervention and to the consolidation behavior observed in the integrated comparison. Conversely, adding the prefix-critic machinery for factor-specific constraint credit on top of CT-Reward yields no detectable incremental change in the reported return or service, resource, and consolidation metrics. This is a non-detection rather than an equivalence result; the present evidence therefore supports common-trace reward credit as the dominant empirical driver of the observed gain after controlling for the factorized surrogate, but does not establish an incremental performance benefit for the prefix machinery. The experiments also do not directly measure policy-gradient variance, so no quantified variance-reduction claim is made.

\subsection{Workload Dependence and Constraint Trade-offs}

The frozen-checkpoint load stress shows that the CT--JC return ordering is not confined to the nominal arrival intensity: the macro-return difference remains positive at all three tested loads, for every load--training-seed pair, and separately under both independent and clustered arrivals. The decomposition is not uniform, however. At low load, the gain is primarily a sensing-resource-efficiency effect and is accompanied by lower completed value and higher positive constraint excess; at high load, completed value shows no detectable CT--JC difference while sensing cost and positive constraint excess both decrease. Robustness here therefore means persistence of the return advantage under the tested intensity shifts, not Pareto dominance of every service and constraint metric. The larger effect at the higher tested load, especially under clustered arrivals, is an empirical trend on this three-point grid rather than a monotonicity or generalization law.

Constraint behavior must likewise be read by family. Hard masks enforce deterministic action validity, whereas sensing-SLA and communication-QoS residuals remain long-term CMDP quantities. In the nominal selected external episodes, CT-PPO has no observed communication-QoS positive excess but modestly higher sensing-SLA excess than JC-PPO, so its lower aggregate $E^+$ reflects a shift across constraint families rather than uniform constraint improvement. Nonzero residuals also preclude interpreting the results as evidence of strict finite-sample feasibility or as a formal safety guarantee.

\subsection{Training and Deployment Implications}

Common-trace credit changes the training procedure rather than the online information set. It requires a simulator or replayable trace generator that can expose stochastic policy replicas to the same exogenous workload realization while their endogenous trajectories diverge. Deployment, by contrast, uses one current public observation, hard masks, the trained Set encoder, and the factorized actor; peer trajectories/returns, prefix values, future arrivals, regime labels, and oracle outcomes are absent. The computational evidence gives the same separation: CT-PPO adds training-side prefix-critic parameters, but the deployed encoder--actor footprint is unchanged. The reference implementation incurs additional latency because it evaluates critics that are unnecessary for action selection, whereas the actor-only path reproduces the reference-path decisions in the equivalence check and has essentially unchanged batch-one CPU latency relative to JC-PPO. These timings are platform-specific and do not imply lower computational complexity or universally identical latency.

\subsection{Scope and Limitations}

The evidence remains limited to the implemented centralized single-cell monostatic ISAC setting with orthogonal sensing/communication resource partitions, equal-share communication scheduling, four discrete sensing profiles, simulator-generated sensing quality, fixed tenant/user/target and task-generation settings, and five training seeds. The load study varies only the arrival-intensity scalars at three values within the same independent/clustered generator families and intentionally reuses matched root identifiers; it is therefore a paired load-shift stress test, not an independent fresh-root replication across loads. Robustness to different tenant/target or task mixes, mobility/channel statistics, sensing-model mismatch, sharing policies, resource budgets, and qualitatively different arrival processes remains untested. Multi-cell/distributed orchestration, adaptive communication scheduling, continuous sensing-resource control, joint waveform/beamforming/precoding optimization, and RF/hardware-in-the-loop or field validation are also outside the present evidence.

\section{Conclusion}
\label{sec:conclusion}

We formulated online multi-tenant sensing-session consolidation as a service-level CMDP and developed CT-PPO, which trains a factorized constrained actor using same-workload stochastic replicas and leave-one-out Monte Carlo return contrasts aligned by physical slot, while retaining factor-specific constraint credit. Replication and prefix critics are training-only; deployment uses the current public observation, hard feasibility masks, Set encoder, and factorized actor.

Across five training seeds and matched external workloads, CT-PPO improves macro return over JC-PPO by $0.934$ (95\% CI $[0.702,1.164]$) and over SLA-Aware Greedy by $1.847$. Relative to JC-PPO, it reduces sensing-resource cost by $6.277$ and increases accepted requests per created session by $0.0806$, linking the gain to selective session reuse. The four-way ablation shows no detectable macro-return improvement from the factor-wise surrogate alone; after controlling for that change, common-trace reward credit improves return by $1.013$ $[0.666,1.338]$ and increases reuse, while the subsequent prefix-critic machinery for factor-specific constraint credit has no detectable incremental return effect. Frozen CT-PPO checkpoints retain a positive return advantage at all three tested arrival loads. CT-PPO's additional parameters are training-side prefix critics; the deployed encoder--actor footprint is unchanged, and measured actor-only batch-one CPU latency remains effectively unchanged. Overall, the evidence links the selective-consolidation gain primarily to common-trace reward credit without increasing the deployed encoder--actor footprint.

\bibliographystyle{IEEEtran}
\bibliography{references}

\end{document}